Bad Weather, Social Network, and Internal Migration; Case of Japanese Sumo Wrestlers 1946-1985.

Eiji Yamamura

**Statements and Declarations**

Competing Interests    There is no conflict of interest to be declared.

Abstract

Post-World War II , there was massive internal migration from rural to urban areas in Japan. The location of Sumo stables was concentrated in Tokyo. Hence, supply of Sumo wrestlers from rural areas to Tokyo was considered as migration. Using a panel dataset covering forty years, specifically 1946-1985, this study investigates how weather conditions and social networks influenced the labor supply of Sumo wrestlers. Major findings are; (1) inclemency of the weather in local areas increased supply of Sumo wrestlers in the period 1946-1965, (2) the effect of the bad weather conditions is greater in the locality where large number of Sumo wrestlers were supplied in the pre-war period, (3) neither the occurrence of bad weather conditions nor their interactions with sumo-wrestlers influenced the supply of Sumo wrestlers in the period 1966-1985. These findings imply that the negative shock of bad weather conditions on agriculture in the rural areas incentivized young individuals to be apprenticed in Sumo stables in Tokyo. Additionally, in such situations, the social networks within Sumo wrestler communities from the same locality are important. However, once the share of workers in agricultural sectors became very low, this mechanism did not work.

1. Introduction

    The farming household population remained almost constant in the pre-World War II period in Japan (Minami, 1967). Under the informal agricultural institution in the pre-war period in Japan, the son was typically designated as the heir to the farmland and had to engage in agriculture. This kept agricultural employment constant until the end of the war (Hayashi and Prescott, 2008). However, the farming household population reached its peak in 1950, and remarkably declined thereafter (Minami, 1967). In the post-war period, miraculous economic growth involved the migration of labor from agriculture to non-agricultural sectors of the economy(Nakamura, 1981).

  A gap in wages was observed to be the key determinant of not only international (Hatton and Williamson, 200) but also internal migration (Kondo and Okubo, 2015; Santiago-Caballero, 2021) as individuals were able to earn more money and receive higher socio-economic status among high-income areas. In Japan, the 1950s were marked by high economic growth and the wage gap caused massive internal migration from agricultural sector in rural areas to non-agricultural sectors in urban areas (Minami, 1967; Mundlak and Strauss, 1978). Crop failures occurred due to unseasonable weather and farmers has to face fluctuations in agricultural output. Hence, weather conditions in the rural areas in Japan should be considered in studying migration patterns. Many recent works have found that weather and climate are key factors in migration (Beine and Jeusette, 2022; Berlemann and Steinhardt, 2017; Cattaneo et al., 2019; Cattaneo and Peri, 2016; Gráda, 2019; Gröger and Zylberberg, 2016; Oliveira and Pereda, 2020; Pajaron and Vasquez, 2020; Trinh et al., 2021)[1]. Extreme heat increases domestic migrations within Mexico, from rural to urban areas (Jessoe et al., 2018). There are various channels through which weather and climate trigger migration. Natural disasters and rainfall shortages result in a widening wage gap between rural and urban areas (Beine and Parsons, 2015). Additionally, natural disasters influence individuals' financial constraints more than their desire to move does (Beine and Parsons, 2017).

  There is an argument that movements out of the agricultural sector were affected by factors other than the wage gap (Ohkawa and Rosovsky, 1973). If labor market information is scarce, moving costs of labor migration decrease with the number of migrants already settled in the destination (Carrington et al., 1996).

---

[1] Labor supply is observed to depend on weather conditions (Ito & Kurosaki, 2009; Jessoe et al., 2018; Krüger & Neugart, 2018; Liu & Hirsch, 2021; Wilson, 2016).

Social networks also play a key role in determining the success of migration (Abad et al., 2021; Beine et al., 2015; Connor, 2019; Goel and Lang, 2019). The seminal work of Munshi (2003) uses weather conditions and social networks to consider the mechanism of migration from developing countries to developed countries. In addition, price of agricultural products depends on climate variability, which used to be larger in the past because agricultural technology was underdeveloped (Ljungqvist et al., 2022). Hence, in Japan, social networks connecting the rural community and the urban areas were more important post World War II

This study examines the role of social networks in promoting internal migration in the post-war period of Japan. For this purpose, we study the labor market for Japanese Sumo wrestlers in the period 1946-1985, covering the process of miraculous economic development[2]. This is because of the novel setting of the Sumo stables which were concentrated in Tokyo, where promising adolescents from all over Japan would apprentice. For employment in the Sumo labor market, migrants from various parts of Japan migrated to the same destination, capital of Japan, Tokyo. Especially after the war, rural agricultural areas are thought to have supplied labor to the sumo wrestling market (Sargeant, 1960). As for wrestlers' educational background, based on the sample of retired wrestlers in 2005, junior high school graduates occupied around 80% of the sample, while higher educated wrestlers were smaller in number before 1989 (Nakajima, 2003)[3]. In 1970, the average age of debut among wrestlers was 16.1 years. Assuming that it takes a year for new entrants to debut(Oinuma, 1994), apprentices in a sumo stable, are thought to have been 15 years old, the age at which compulsory education is completed. This is consistent with the fact that most applicants entered the Sumo stable directly after completing basic compulsory education (West, 1997). This means that Sumo wrestlers decided to be apprenticed in a sumo stable before graduating from school. If weather and migration are closely related, decision-making would depend on the climate conditions prevalent one year before the completion of education of the apprenticed sumo wrestlers, considering that they are likely to be farmers' heirs.

In the present day, it is difficult to obtain information about Sumo stables. Especially for less educated rural adolescents, who had less choice regarding their Sumo stable apprenticeship preferences. In residential and communal areas, several community members would have connections to Sumo stables where wrestlers who originated from their locality worked The social network connections between the community and the Sumo stables in Tokyo are stronger and larger because the community produced lager number of sumo wrestlers, enhancing migration to the Sumo labor market. Analyzing the panel data of new entrants in Sumo stables, we found that the network plays a greater role in supplying new entrants, especially when bad weather conditions occurred in the developing stage. However, this tendency is not observed in the developed stage when share of agriculture in the national GDP, became less significant. This study provides the evidence that bad weather conditions and social networks influenced migration from rural areas to urban areas in the post-war period, substantially hampering the long-term development process.

---

[2] In the globalization era, international immigration had a great impact on Sumo wrestling (Yamamura, 2014). In contrast, the Sumo labor market shrank in the studied period.
[3] In the Japanese educational, compulsory education is completed at junior high school, usually when students become 15 years old.

The rest of the paper is organized as follows: Section 2 overviews the post-war setting Japan. Section 3 explains the dataset. In Table 4, simple econometric framework is provided. Section 5 reports the results of the estimations and its interpretation. The final section offers concluding observations.

2. Overview of the Sumo wrestling realities of post-war period in Japan.

After World War II , Japan experienced miraculous economic growth and a drastic change in the life style. Considering the entire process of the economic development, this study put focus on the period ranging 1946-1985. Immediately after the war, Japan was in the developing stage and leisure-oriented goods and services were scarce. In the 1950s, urban areas such as Tokyo, Osaka, and Nagoya, witnessed more participation in western sports, such as baseball, compared to participation ancient and traditional sports, such as Sumo. In many parts of the country, particularly in farming villages, a contrasting picture was found (Sargeant, 1960). The economic reason for Sumo's popularity within adolescent males in farming villages is described as "the farmers in the north eke out a bare living, the land is poor, crop failures are not uncommon. They have little money to spend on their sons' pastimes. Consequently, the boys take up the cheapest sport they can find. Sumo, of course, fills the bill perfectly" (Sargeant 1960, p.12).

Typically, sumo stable scouts recruited males who were in their mid-teens (15-17 years old). Most recruits had only the most basic and compulsory junior high school education (West, 1997). Less educated adolescents left their hometowns and were apprenticed to Sumo stables in Tokyo. In the post-war period, rail-way expanded all over Japan and internal migration was promoted because of low transportation costs (Sequeira et al., 2020). However, because of the lack of information regarding Sumo stables, adolescent males, especially undergraduate applicants, were incapable of choosing Sumo stables for themselves. According to the anecdotal episode of a wrestler, the advice of Sumo leaders from their hometown was useful in choosing a stable to apprentice in, because the leaders had connections with certain stables in Tokyo (Miyabiyama, 2014). Hence, promising young adolescents entered Sumo stables through the connections of local leaders, which further strengthened the network. Consequently, Sumo stables received new entrants mainly from certain areas. Consistent with previous studies considering role of networks in migration(Abad and Sánchez-Alonso, 2018; Goel and Lang, 2019; Munshi, 2003), applicants that are able to enjoy the benefits of the network, enter Sumo stable when the networks are stronger.

The real situation of the Japanese society changed drastically because of high economic growth, placing Japan among the developed countries of the world. Symbolically, in 1964, Tokyo Olympics were held, recording the first instance of the Olympics being held in a non-Western country. The 1964 Tokyo Olympics made it evident that Japan had recovered from the war and received acknowledgment from the whole world. Therefore, in this study, the first period covers the 20 years from 1946 to 1965.

After the 1964 Tokyo Olympics, Japan is considered to have entered the developed stage, where Japanese life style was distinctly differed from the previous stage. In this stage, the contribution of agricultural sectors to GDP was smaller than ever before. Hence, labor supply from agricultural areas to Sumo wrestling was not considered to be large. The emergence of the first foreign-born Sumo wrestler in 1968 reflected the situation which is regarded as a turning point in the history of Sumo wrestling. Inevitably, mechanisms of labor supply from rural agricultural areas would change. However, from mid-1985, new entrants from abroad increased remarkably and Sumo wrestling entered its globalization period (Yamamura, 2014). The purpose of this study is to examine this labor supply, mainly in the domestic labor market. Hence, the second period covers the 20 years from 1965 to 1985, before full-fledged globalization took effect.

3. Dataset

The data covers localities throughout Japan in the period 1946-1985. The key variable is the number of new entrants to Sumo stables from the localities. As units of locality, we use prefectures, which are the basic local entities of Japan and prefecture-level data is more easily available than data from any other local entities. There are 47 prefectures in the present day, but Okinawa prefecture was not included in Japan from 1946 to 1972, as it was under the rule of the United States. Hence, the data set covered 46 prefectures. The panel data is constructed from various sources explained as follows.

Dataset about Sumo Wrestlers such as their debut ages and birthplaces are obtained from various sources (Mizuno & Kyosu, 2011; Sumo editorial staff (eds.), 2001), and website database[4]. In these database, accurate individual-level data of Sumo wrestlers is available from 1968, in the period of the Meiji Restoration. In the studied period (1946-1985), the total number of new entrants is recorded at 3,723. Following the argument presented in Section 2, we divide the forty-year periods into two halves, namely pre- and post-Tokyo Olympics, 1964. Accordingly, new entrants are recorded at 1,397 and 2,326 in the periods ranging 1946-1965 and 1966-1985, respectively. To put it more precisely, Figure 1 demonstrates that the number of new entrants recorded is below 50 nearly ten years after the World War II. This number is then observed to have increased drastically and reached over 100 recorded entrants in 1958, after which it stays over 100 in most years. GDP and rate of production in agricultural sectors in each prefecture are available from the government's official data, although these data are available only onwards 1955[5].

Figure 2 illustrates the consistent decline in rate of agricultural production from 1955 to 1985. Considering Figures 1 and 2, it is jointly suggested that rate of growth of non-agricultural sectors is positively related to the number of new entrants in Sumo wrestling. Sumo wrestling is a part of the non-agricultural sector, therefore labor from agricultural sectors moved to Sumo wrestling. However, an increase in number of new entrants is not observed after the 1960s, whereas a decreasing trend in the rate of agricultural production persists throughout the period. One possible interpretation is that jobs in non-agricultural sectors varied and increased with economic growth. Further, there were more well-educated young people in this period as compared to the 1950s. Therefore, they were able to find jobs that were more promising than sumo wrestling. The shrinking agricultural sector naturally reduced the main source of labor

---

[4] Website of "Sumo Reference". http://sumodb.sumogames.de/Rikishi.aspx?shikona=&heya=-1&shusshin=-1&b=-1&high=-1&hd=-1&entry=-1&intai=-1&sort=1&l=e. Accessed on 26 October, 2015.

[5] Website of Cabinet Office, Government of Japan http://www.esri.cao.go.jp/jp/sna/data/data_list/kenmin/files/contents/main_68sna_s30.html. Accessed on Oct 28, 2015

supply for Sumo wrestling, leading to a decrease in new entrants. Naturally, Sumo recruiters started searching for talent foreign countries. Consequently, foreign wrestlers entered the Sumo wrestling scene in 1968 (Yamamura, 2014).

Historical data about natural disasters are available from National Astronomical Observatory of Japan (Shizen Kagaku Kenkyu Kiko, 2016). The data also shows whether natural disasters had a damaging impact on agriculture. There are various types of disasters, and some types of disasters did not influence agricultural production. The damage was mainly caused by harsh weather conditions such as cold temperatures and typhoons with heavy rains. Actually, migration increased after such climate variations were observed, but not after other types of natural disasters occurred (Bohra-Mishra et al., 2014)[6]. Therefore, we studied the effect of the disasters that inflicted damage on agricultural production. Further, the data also indicates the years of the disasters and the regions which suffered from the disasters. Hence, we integrated the information regarding disasters with data on Sumo wrestlers. However, there is no information about the varying degree of damages among regions, though there was difference in degrees of damage among the stricken prefectures. Due to the limitations of the data, the dummy variable takes the value of 1 only if the agricultural production was damaged by the disaster, otherwise, it takes 0. Figure 3 demonstrates number of damaged prefectures during the studied period. In Figure 3, it is evident that there is wide variation without regularity indicating the unpredictability of the disasters

Based on the dataset which is used for Figure 1, Figures 4(a) and (b) are illustrated. In each Figure, vertical and horizontal lines indicate the number of new entrants from each prefecture in the later period and the earlier period, respectively. In both Figures 4(a) and (b), a positive relation is observed between them, which means that prefectures that supplied larger number of Sumo entrants in the earlier period, continued to supply a large number of Sumo entrants in the later period. This indicates that social network leaders in certain prefectures guided the new entrants to a certain degree. Otherwise, applicants obtained information on Sumo stables from incumbent wrestlers. This suggests that the social networks between rural areas and Sumo stables in urban areas, play a role in the labor supply of Sumo wrestlers.

In order to capture the importance of agriculture in the Japanese economy, rate of agricultural product over GDP was gathered. This study considers economic conditions prevalent not only in the post-war period, but also the pre-war period, to investigate the persistence of pre-war conditions. The rate of agricultural sector production is unavailable in the official statistics for the pre-war period. Accordingly, data about rate of workers engaged in the agricultural sector in 1940, are used. This dataset was provided by the Research Unit for Statistical and Empirical Analysis in Social Sciences in Hitotsubashi University[7].

---

[6] No influence of eruption, landslide, and earthquake has been observed.

The description and basic statistics of variables used in this study are exhibited in Table 1. As for *Entrants*, the average number of new entrants are 1.48, while maximum and minimum values are 23 and 0. Therefore, there is large gap in number of new entrants between prefectures and its values are skewed towards the left. The studied period is limited to the post-war period. However, there is a possibility that the social network connecting rural areas with Sumo stables in Tokyo was formed through long-term interactions, which persisted even after the war ended. On this assumption, percentage of new entrants in each prefecture over the total number of new entrants, is calculated for the pre-war and (1868-1945) post-war (1946-1965) periods. Mean value of *Rat Entrants 1868-1945* is 2.17, which when multiplied by 46, is 98. This is reasonable to show small difference from 100 (%) because of round error. There is wide variation because minimum and maximum values *Rat Entrants 1868-1945* are 0.13 and 19.2. Basic statistics of *Rat Entrants 1946-1965* shows the similar values to *Rat Entrants 1868-1945*. Minimum value of *Rat Entrants 1946-1965* is 0 because there was nobody entering the Sumo stable from Wakayama prefecture from 1946 to 1965. *Distance* is measure as the distance between Tokyo and the capital of each prefecture[8]. Therefore, minimum value of *Distance* is 0 for Tokyo.

4. Econometric framework

In Japan, one year before their graduation, students face the following choices in a given year *t*: local agricultural work, local non-agricultural work, migration or higher level of education (Jessoe et al., 2018). The conditions changed according to the development stages. In the developing stage, probability of pursuing higher level of education and the probability to work in local non-agricultural sector, was very low. In actuality, there were only two choices available; local agricultural work or migration. Observing the weather condition in the *t-1* period, it was possible to predict the probability of employment in local agricultural sector in the next year, *t*. Bad weather conditions reduce the probability to work in local agricultural sector and motivate migration (Jessoe et al., 2018; Munshi, 2003). In this stage, weather fluctuations have a greater effect on outcome as compared to the more developed stages, where more choices are available. In short, bad weather conditions in the *t-1* period increase migration in the *t* period. Further, in the model of costly search, job seekers are benefitted when the employed members of their network inform them about newly available jobs (Carrington et al., 1996). In order to examine how weather conditions and social networks influence labor supply for Sumo wrestling, the form of the baseline estimated function is:

$$Entrants_{it} = \alpha_1 Bad\ weather_{it-1} + \alpha_2 Past\ Rat\ Entrants_i + \alpha_3 Agriculture_{it} + \alpha_4 Distance_i + \alpha_5 Population_{it} + m_t + u_{it} \quad (1)$$

where *Entrants $_{it}$* represents the dependent variable for prefecture *i* for the year *t*. When *Entrants* is counted as a variable, Poisson model is used (Wooldridge, 2002). The regression parameters are denoted as *α* and the error term is $u_{it}$. $m_t$ captures effects of time periods which are controlled by year dummies. Hence, the effects of macro-level economic shocks on all parts of Japan are controlled.

---

[7] Details of the estimated data are explained in the report of Prof. Fukao team (Yuan et al., 2009).
[8] Website of Geographical Information Authority of Japan. https://www.gsi.go.jp/KOKUJYOHO/kenchokan.html. Accessed on 30 October, 2015.

From the previously presented argument, bad weather conditions are assumed to influence decision-making and job searching in the next year (Jessoe et al., 2018; Munshi, 2003). Hence, we include the variable *Bad weather* for the t-1 period. Coefficient of *Bad weather* would indicate a positive co-efficient if bad weather increases the number of entrants in Sumostables.

The effect of the social networks between locality and Sumo stables is considered to be stronger if the number of new entrants in the past is larger (Carrington et al., 1996). *Past Rat Entrants* is included to capture the effect of the social network. The expected sign of coefficient for past entrants is positive. There are two reasons that number of new entrants in the previous years is not used to capture the network effect. First, endogenous bias occurs if number of new entrants in the previous year is used. Therefore, it is necessary to consider sufficient time lag when information regarding past entrants is used. Second, the social network was formed and preserved through repeated interactions over time. Hence, in this regard, social network is not considered to vary temporarily in accordance with the situations of last year. As explained in the previous section, data of Sumo wrestlers is available from 1868 and so it is valuable to use the data as old as possible. Therefore, data from 1868 to 1945 are used to quantify the network effects between rural areas and Sumo stables in Tokyo, in the pre-war period. However, data from 1946 to 1965 can be used to quantify social networks when network effects in the period 1966-1985 are examined. Therefore, in the estimation, two variables, *Rat Entrants 1868-1945* and *Rat Entrants 1946-1965,* are used as proxy for social networks in sub-samples where data from the first and second period are used. Therefore, to consider long-term network effects, *Rat Entrants 1868-1945* are used in estimations using both sub-samples. *Rat Entrants 1946-1965* is used when a sub-sample covers the period 1966-1985, to examine the more recent network effects.

Considering the control variables, effects of *Agriculture* are captured by two variables. Before 1955, due to lack of relevant data, percentage of workers in agricultural sector in 1940 is used for the period 1946-1965. The percentage of production in agricultural sector is used but sample size was reduced because the sub-sample covers only the period 1955-1965. Similarly, prefecture-level GDP is available from 1955, and *GDP* is included when the sub-sample covering the period 1955-1965 is used. Apart from this, population size is controlled by incorporating *Population*.

$$Entrants_{it} = \beta_1 Bad\ weather_{it-1} + \beta_2 Bad\ weather_{it-1} * Past\ Rat\ Entrants_i$$
$$+ \beta_3 Agriculture_{it} * Past\ Rat\ Entrants_i + \beta_4 Agriculture_{it} + \beta_5 Population_{it}$$
$$+ k_i + m_t + u_{it}, \qquad (2)$$

Additionally, unobservable prefectures' time-invariant features, represented as $k_i$, are controlled by the Fixed Effects Poisson model (Wooldridge, 2002). Hence, *Distant* and *Past Rat Entrants* are constant in different years and completely controlled by the Fixed Effects, because the controlled model is equivalent to the model with 46 prefecture dummies. Consequently, *Distant* and *Past Rat Entrants* are automatically excluded from the model.

Different from the baseline model, this model adds *Bad weather * Rat Entrants 1868-1945* (or *Bad weather *Rat Entrants 1946-1965)* as a key variable. It is the cross term between *Bad weather* and *Rat Entrants 1868-1945* (or *Rat Entrants 1946-1965)*. Whether *Rat Entrants 1868-1945* is included depends on the period of the sub-sample and what we tend to examine in the specification. Even in the estimation during the period 1965-1985, *Bad weather * Rat Entrants 1868-1945 is* included if we examine whether the long persistent effect of the social network functioned under the difficult situation caused by the bad weather. The coefficient should be positive if role of social network is more important when the bad weather conditions occur and reduce the probability of to work in the hometown. Apart from this, bad weather conditions show greater damage to the areas where agriculture is more important for the local economy. In order to control this effect, we add *Bad weather * Agriculture* as a cross term between *Bad weather* and *Agriculture*. Its expected co-efficient should be positive because negative effects of bad weather conditions on local economy are greater in agricultural areas, leading to increased migration.

These specifications differ slightly in accordance with the sub-sample used for estimations because of data availability. In the later period, most variables in the model are available for every year and can be included. For instance, *Agriculture* and *GDP* are included in the model for the period 1966-1985, but not for the period 1946-1965. Estimation results are also comparable between periods to examine how the importance of social networks and impact of bad weather conditions changed over time, as labor supply to sumo stables is explored.

5. Estimation results and their interpretation

Tables 2 and 3 report results using the sub-samples for 1946-1965 and 1955-1965, respectively, while Tables 4 and 5 exhibit the results using the sub-sample for 1966-1985 only. In these Tables, column (1) presents the results of Poisson regression as baseline specifications. The results of the Fixed Effects Poisson regressions are shown in other columns where cross terms are included as key variables of this study.

In column (1) of Table 2, a positive sign for coefficients of *Bad weather* with statistical significance at the 1 % level is observed, which indicates the inference that occurrence of bad weather conditions cause an increase in labor migration of Sumo wrestlers in the next year. Absolute value of the coefficient is 0.90, which indicates that new entrants increased by a magnitude of 0.90 in the years following    bad weather conditions. This is consistent with existing works(Bohra-Mishra et al., 2014; Jessoe et al., 2018; Mueller et al., 2014). The significantly positive co-efficient of *Rat Entrants 1868-1964* indicates that labor supply for Sumo wrestling is greater for prefectures from which larger number of wrestlers historically migrated. This indicates that new entrants search Sumo stables through social networks formed in the past, because potential recruits can benefit from Sumo stable-related information that they receive from incumbent wrestlers or middleman. The absolute value of coefficient of *Rat Entrants 1868-1964* is 0.06. This means that 1 % higher share of wrestlers coming from the area in the past leads increase in new entrants by 0.06. As shown in Table 1, mean value of new entrants is 1.48. Therefore, these values imply a 10 % increase in the share of past wrestlers, resulting in a 40 % increase in new entrants[9]. Therefore, effect of the network on new entrants is sizable.

*Agriculture 1940* produces a positive co-efficient and is statistically significant at the 1 % level, meaning that Sumo wrestlers were likely to migrate from agricultural areas. The significantly positive co-efficient of *Population* suggests that more talented and promising recruits exist in populated areas, if other things are considered constant. Surprisingly, *Distance* indicates a significantly positive co-efficient, which implies that new entrants are more likely to migrate from distant areas. One possible interpretation is that; recruits from rural and agricultural areas distant to Tokyo, were more motivated to be apprenticed in Sumo stables.

Concerning the results of the Fixed Effects Poisson estimation described in columns (2)-(5), *Bad weather* and *Population* variables indicate similar results in column (1), although other control variables are not included because these variables were completely controlled by the model. The absolute values of coefficient of *Bad weather* and *Population* are 0.90 and 0.39 as shown in columns (1) and (2), respectively. The Poisson Regression result suffers from upward biases because unobserved time-invariant features of prefectures are not controlled. Therefore, the unbiased effect is observed at 0.39, implying that new entrants increased by 0.4 points in the years following bad weather conditions. Table 1 indicates that mean value of new entrants is 1.48. Therefore, occurrence of bad weather conditions increases new entrants by approximately 35 %.

Turning to columns (3)-(4) in Table 2, the variable *Bad weather \* Rat Entrants 1868_45* indicates a positive co-efficient with statistical significance. Consistent with the previous inference, social networks are more effective in supplying labor to Sumo stables, especially when farmers face difficulty due to bad weather conditions. Therefore, the negative impact of bad weather conditions on local labor market was mitigated by connecting local and urban labor markets through social networks. In column (4), the absolute values of the coefficient of *Bad weather \* Rat Entrants 1868_45* is 0.03, implying that network effect is larger by 0.03 points in years following bad weather conditions. As interpreted above, value of the coefficient of *Rat Entrants 1868_45* is 0.06 in column (1). Therefore, social network effect increases by 50% one year after bad weather conditions were faced. In contrast, *Bad weather * Agriculture* did not show statistical significance in columns (3) and (5). Therefore, effect of the share of agriculture in the area on the labor supply for Sumo stables did not depend on weather conditions. In our interpretation, even when the agricultural area suffered bad weather conditions, potential recruits were not able to search for Sumo stables to apprentice in, if they did not have the opportunity to be acquainted with middlemen or incumbent wrestlers. In the period 1946-1965, the search cost for uneducated persons looking to enter Sumo wrestling was large for the rural areas.

---

[9] (0.6/1.48) × 100=40.1 %

Table 3 reports the results where *GDP* and *Agriculture* are included to mitigate the omitted variables biases and we can use the sub-sample of the period when these variables are available. As a whole, results of Table 3 are similar to those of Table 2. Especially, values of coefficients of *Bad weather * Rat Entrants 1868_45 Rat Entrants 1868_45* are 0.03 in columns (3) and (4), which are the same as those in Table 2. Therefore, in the economic developing stage, we observed robust results indicating that social networks play a crucial role in mitigating the difficulty caused by the bad weather conditions.

Moving forward, we examine the trends of the developed stage, during the period 1966-1985. Specifications of Table 4 are almost equivalent and comparable to Table 2, although we included different variables to capture the share of agriculture in the area. In Table 4, key variables such as *Bad weather* and *Rat Entrants 1868_45* indicate results which are different from those in Table 2. *Bad weather* is not statistically significant while *Rat Entrants 1868_45* presents a significantly negative co-efficient. One explanation about the negative co-efficient of *Rat Entrants 1868_45* is that structural changes in Japanese economy lead the social network formed before the end of the war to have opposite effects. For instance, the transportation system had developed drastically since 1945. This lead to an increase in the number of new entrants from areas where wrestlers had hardly migrated before 1945. That is, the network persists in the developed stage after the War, but not in the developed stage. Further, cross terms do not show statistical significance in any columns. Meanwhile, control variables such as *Distance, Agriculture* and *Population,* are observed to have significant positive signs, which are similar to those of Table 2.

In Table 5, instead of *Rat Entrants 1868_45*, alternative social network variable, *Rat Entrants 1946_65,* is used. In column (1), the significant positive sign of *Rat Entrants 1946_65* is observed, which implies the existence of social network and its critical role in the developed stage, after Japan experienced economic growth. Hence, search costs remain very high due to lack of Sumo labor market information. Considering the results of the network in Tables 4 and 5 together, we derive the argument that newly formed social networks becomes substantially more important than old networks, in the process of economic development. Concerning key variables, *Bad Weather* and *Bad weather \* Rat Entrants 1946_65* do not show a significantly positive co-efficient, which is inconsistent with   Tables 2 and 3. We interpret it as follows. The share of agriculture sector drastically declined, even in the rural areas. This reduced the labor supply to Sumo stables, and so bad weather conditions hardly influenced the labor supply.

   The observations made in this study made it evident that social networks were effective in facilitating internal migration, from rural to urban areas for Sumo wrestling during the period 1946-1965. The role of social networks is especially important when rural areas suffer from bad weather conditions. The mechanism is similar to that of international immigration from developing countries to developed countries (Munshi, 2003). However, after entering the developed period, period in which the share of agricultural production declined drastically, social networks remained important but and their effect were   not influenced by weather fluctuations. In comparison to the period directly after World War II, most people entered higher levels of education after completing compulsory education and were able to find jobs in non-agricultural sectors, not only in urban but also in rural areas. Subsequently, students face various choices before their graduation. Hence, decision-making about the choice of jobs become less likely to depend on fluctuations in weather in the developed stage, which is reflected in this study. The contribution of this study is to provide evidence that role of social networks has changed in the process of economic development, using historical record of internal migration in Japanese Sumo labor market.

6.  Conclusion

   This paper is the first contribution based on a historical perspective to unfold the interaction between weather conditions and the social networks when migration mechanism is investigated in labor market of Japanese Sumo wrestling, Massive internal migration from rural to urban areas was observed in the developing countries. Recently, researchers paid much attention to the fact that weather fluctuations influence migration from rural agricultural areas to urban areas. However, the impacts of weather conditions and changes in the social networks have not been sufficiently investigated. Hence, this study aims to bridge economic history and modern development economics. Historical data related to the occurrence bad weather conditions is matched with number of new entrants to Sumo stables from each prefecture in each year studied. Further, based on the number of wrestlers from each prefecture in the pre-war period, we have calculated the percentage of each prefecture's wrestlers among all wrestlers. This was used as a proxy variable for examining community networks and was matched with the panel data. Major findings of the Fixed-Effects Poisson regression are; (1) Extremely cold weather in local areas increased supply of sumo-wrestlers in the next year in period 1946-1965, (2) the effect of cold weather is larger in the locality where social networks were stronger in the pre-war period, (3) neither occurrence of bad weather nor its interaction with social networks in the past influenced the labor supply in the period 1966-1985.

Findings of this study made it evident that, in early stage of economic development, school students in rural agrarian areas decided to leave for the capital, Tokyo, and apprenticed in Sumo stable through community-based social networks. The network effect is more effective if bad weather conditions were observed in the one year before students completed their schoolings. The migration mechanism is similar to the one observed in Munshi (2003). However, this mechanism disappeared in the more developed stage, partly because non-agricultural sectors occupied most of the labor population, and social networks were less likely to function. School students were able to find jobs in a variety of markets where their expected income was higher than that earned from Sumo wrestling. Labor supply in the Sumo labor market drastically changed during the period 1946-1985.

The interactions between weather conditions and social networks for migration purposes, has important implications that are required to be further explored. Due to limitations of the historical dataset, we could not obtain detailed information about how bad weather conditions affected the agricultural yield in each prefecture. Further, it is unknown whether new entrants were reared on farms. We could identify who suffered income shock because of bad weather conditions if we could have obtained the detailed data. The extent of reduction in agricultural yield caused by bad weather conditions should be explored in the context of migration in future studies. Furthermore, the role of inter-personal networks in non-agricultural sectors is valuable to study the developed stage discussed in this study. That is, future research should thoroughly investigate how interactions between fluctuations in income and inter-personal networks affect labor supply, on a long-term historical basis.

Figure 1 Total number of new entrants to Sumo stables

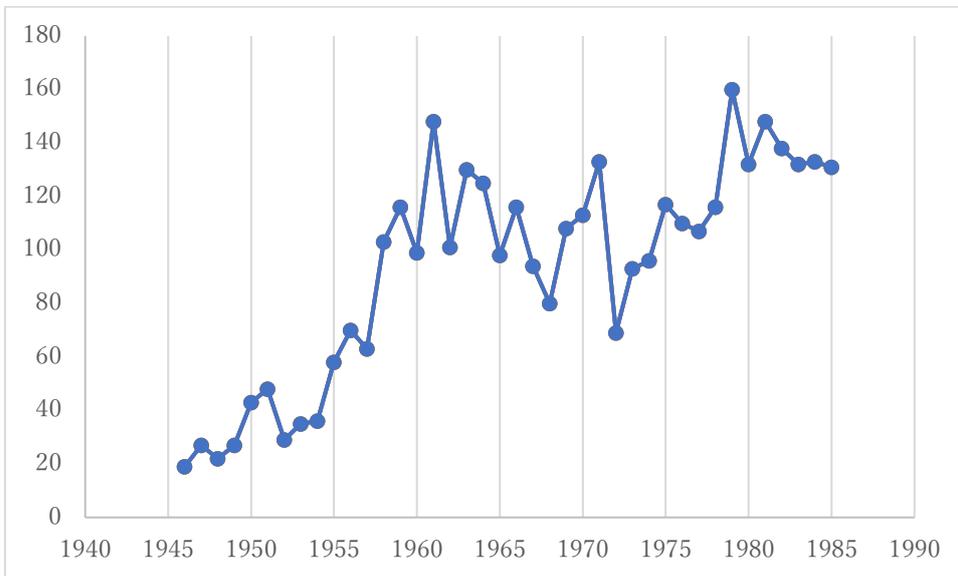

Source: Sumo References

http://sumodb.sumogames.de/Rikishi.aspx?shikona=&heya=-1&shusshin=-1&b=-1&high=-1&hd=-1&entry=-1&intai=-1&sort=1&l=e. Accessed on 26 October, 2015.

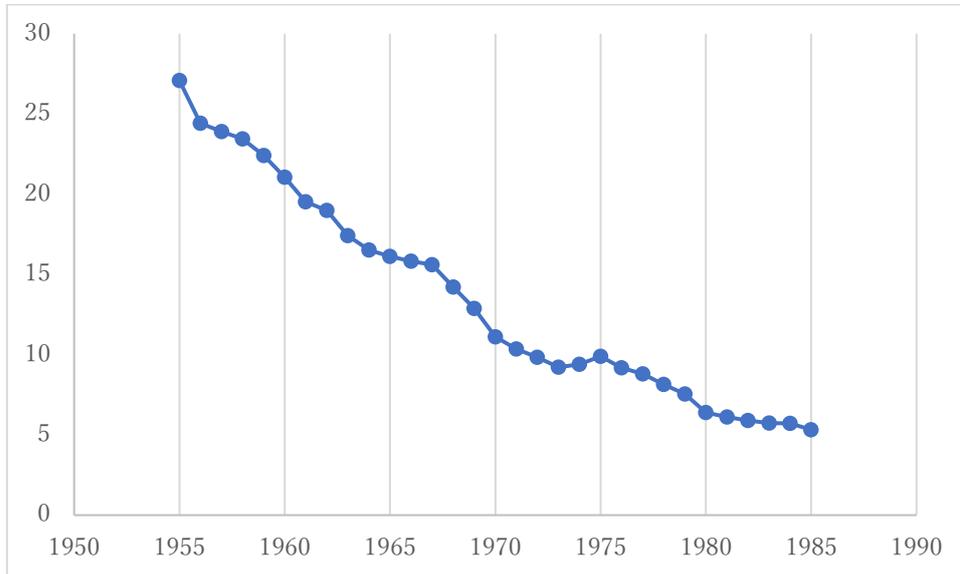

Figure 2 Rate of agricultural production (%)

Source: Website of Cabinet Office, Government of Japan
http://www.esri.cao.go.jp/jp/sna/data/data_list/kenmin/files/contents/main_68sna_s30.html

Figure 3. Number of prefectures that suffered from cold temperatures, damaging agricultural production.

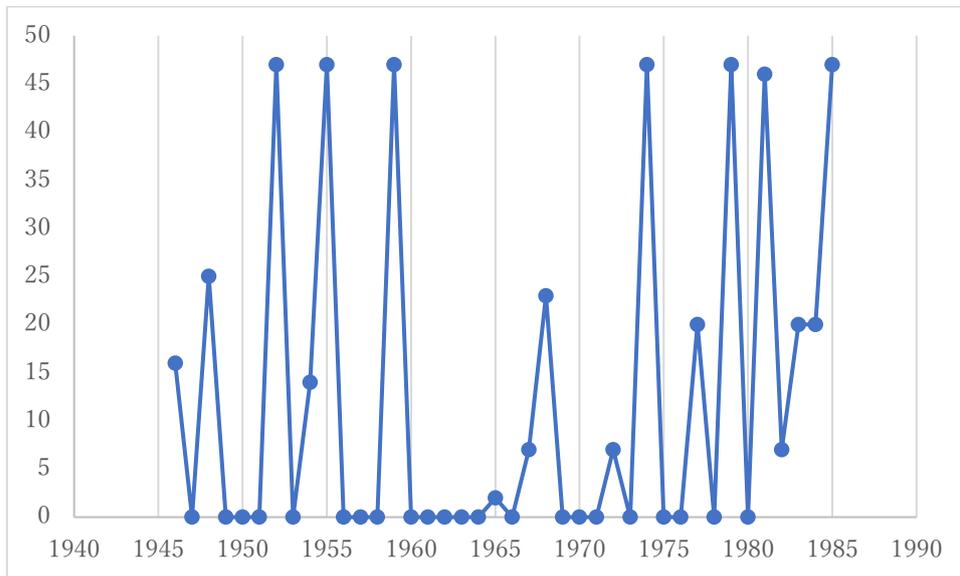

Note: Bad weather occurred in a past year. However, it affected the labor market in the next year. Therefore, its values are shown in the year succeeding the year of occurrence.

Source: Shizen Kagaku Kenkyu Kiko, K. T. (eds.) (2016). Rika Nenpyo 2016 [Chronological Scientific Tables 2016]. Maruzen Publishing.

Figure 4(a). Relation between number of sumo wrestlers in the period (1868_1945) and their numbers in the post-war period (1946-1965). (%)

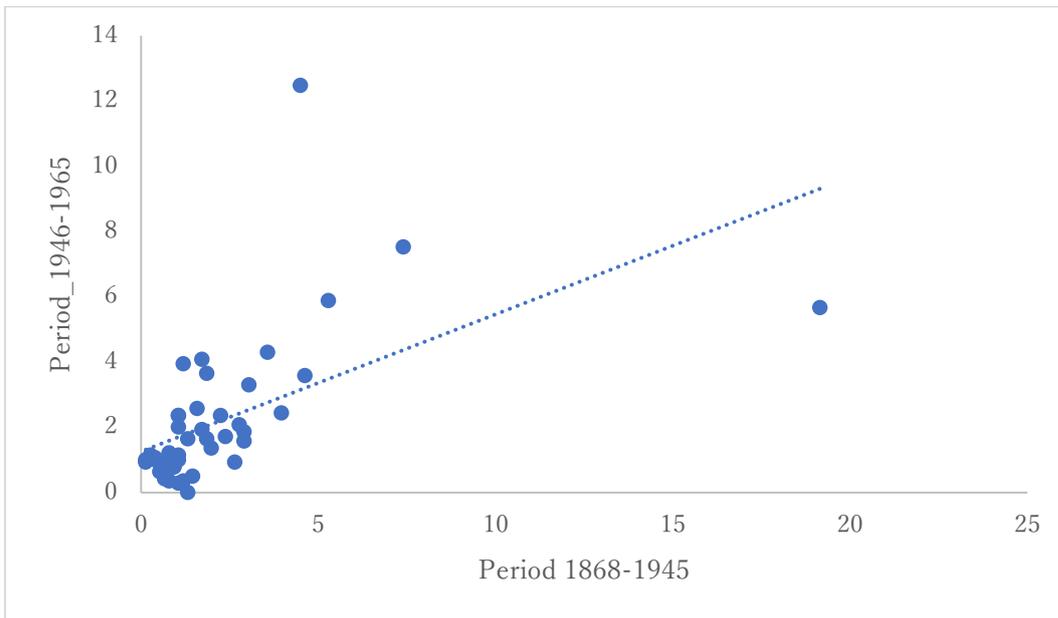

Figure 4(b). Relation between number of sumo wrestlers 1946-1965 period and their numbers in the period (1966-1985). (%)

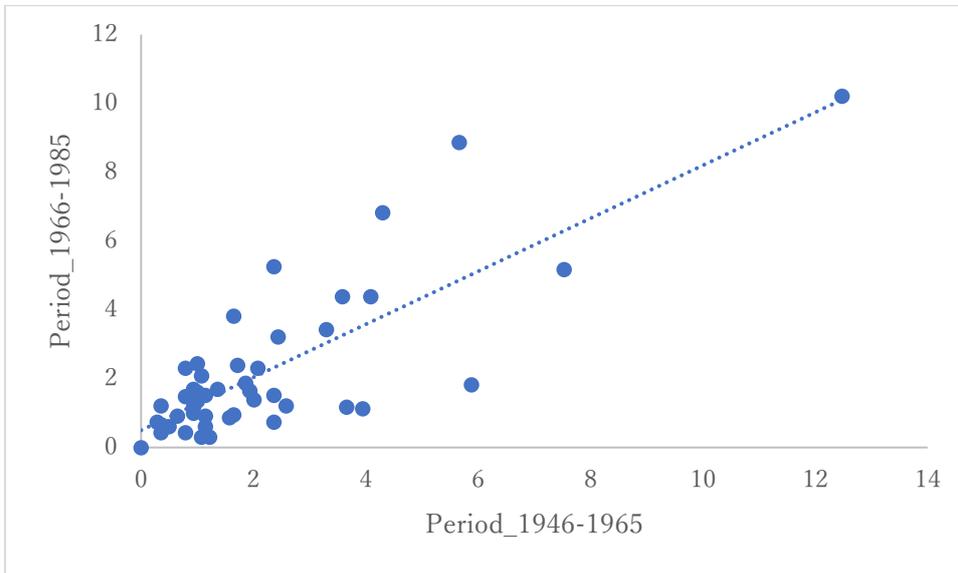

Source: Sumo References

http://sumodb.sumogames.de/Rikishi.aspx?shikona=&heya=-1&shusshin=-1&b=-1&high=-1&hd=-1&entry=-1&intai=-1&sort=1&l=e. Accessed on 26 October, 2015.

Table 1. Basic statistics of variables used in the estimation

| Variables | Definition | Mean | Max | Min |
|---|---|---|---|---|
| *Entrants* [a] | Number of entrants into Sumo stables | 1.48 | 23 | 0 |
| *Rat Entrants 1868_45* [a] | Rate of entrants into Sumo stables in the period 1868-1945 (Entrants in pref/ total entrants: %). | 2.16 | 19.2 | 0.13 |
| *Rat Entrants 1946_65* [a] | Rate of entrants into Sumo stables in post-war period 1946-1965 (Entrants in pref/ total entrants: %). | 2.16 | 12.5 | 0 |
| *Bad weather* [b] | Occurrence of bad weather conditions in previous year | 0.21 | 1 | 0 |
| *Distance* [c] | Distance between Tokyo and home prefecture (1000km) | 0.45 | 1.55 | 0 |
| *Population* [d] | Population (10 million) | 1.91 | 10.8 | 0.55 |
| *Agriculture 1940* [e] | Rate of workers in agricultural sector in 1940 (%) | 24.1 | 33.5 | 1.96 |
| *Agriculture* [f] | Rate of production in agricultural sector (%) | 20.1 | 43.8 | 0.45 |
| *GDP* [f] | GDP per capita (million yen) (available from 1955) | 0.15 | 0.52 | 0.05 |

Sources:

a. Sumo References

http://sumodb.sumogames.de/Rikishi.aspx?shikona=&heya=-1&shusshin=-1&b=-1&high=-1&hd=-1&entry=-1&intai=-1&sort=1&l=e. Accessed on 26 October, 2015.

b. Shizen Kagaku Kenkyu Kiko, K. T. (eds.) (2016). Rika Nenpyo 2016 [Chronological Scientific Tables 2016]. Maruzen Publishing.

c. Website of Geographical Information Authority of Japan.

https://www.gsi.go.jp/KOKUJYOHO/kenchokan.html. Accessed on 30 October, 2015.

d. Statistics Bureau

   http://www.stat.go.jp/data/chouki/02.htm. accessed on Oct 28, 2015

e. Hi-Stat Social Science Database Network, which is accessed through website of Institute of Economic Research, Hitotsubashi University.

http://www.ier.hit-u.ac.jp/Japanese/databases/#09. Accessed on 10 September, 2017.

f. Website of Cabinet Office, Government of Japan

http://www.esri.cao.go.jp/jp/sna/data/data_list/kenmin/files/contents/main_68sna_s30.html.

Accessed on Oct 28, 2015

Table 2. Determinants of number of entrants in Sum stables For period 1946-1965: Full sample

|  | (1) Poisson | (2) FE Pois | (3) FE Pois | (4) FE Pois | (5) FE Pois |
|---|---|---|---|---|---|
| *Bad weather* | 0.90*** (0.17) | 0.39*** (0.13) | 0.25 (0.25) | 0.28* (0.16) | 0.51** (0.25) |
| *Rat Entrants 1868_45* | 0.06*** (0.02) | | | | |
| *Bad weather* RatEntrants 1868_45* | | | 0.03* (0.02) | 0.03*** (0.01) | |
| *Bad weather* Agriculture 1940* | | | 0.001 (0.009) | | −0.005 (0.009) |
| *Agriculture 1940* | 0.03*** (0.01) | | | | |
| *Distance* | 1.55*** (0.12) | | | | |
| *Population* | 0.25*** (0.05) | 0.32*** (0.05) | 0.36*** (0.05) | 0.36*** (0.05) | 0.34*** (0.04) |
| Observations | 920 | 900 | 900 | 900 | 900 |
| Group | 46 | 45 | 45 | 45 | 45 |
| Wald stat | 636.2 | 1781.1 | 2656.8 | 2089.8 | 2655.9 |

Note: Year dummies are included in all estimations although its results are not reported. Numbers in parentheses are robust standard errors clustered at prefecture. *, **, and *** indicate significance at the 10%, 5%, and 1% levels.

Table 3. Determinants of number of entrants in Sumo stables for the period 1946-1965: Sub-sample with GDP

| | (1) Poisson | (2) FE Pois | (3) FE Pois | (4) FE Pois | (5) FE Pois |
|---|---|---|---|---|---|
| *Bad weather* | 0.99*** | 0.39** | 0.15 | 0.24 | 0.39 |
| | (0.20) | (0.18) | (0.31) | (0.19) | (0.25) |
| *Rat Entrants 1868_45* | 0.004 | | | | |
| | (0.02) | | | | |
| *Bad weather* * RatEntrants 1868_45* | | | 0.03* | 0.03** | |
| | | | (0.02) | (0.02) | |
| *Bad weather* * Agriculture* | | | 0.003 | | 0.0001 |
| | | | (0.008) | | (0.01) |
| *Distance* | 1.07*** | | | | |
| | (0.14) | | | | |
| *Agriculture* | 0.05*** | −0.0003 | −0.0002 | 0.001 | −0.0003 |
| | (0.07) | (0.02) | (0.02) | (0.02) | (0.02) |
| *Population* | 0.40*** | −0.25 | −0.18 | −0.19 | −0.26 |
| | (0.06) | (0.23) | (0.24) | (0.24) | (0.23) |
| *GDP* | −0.97 | 7.17** | 6.88** | 7.14** | 7.16** |
| | (1.79) | (3.29) | (3.35) | (3.24) | (3.38) |
| Observations | 506 | 495 | 495 | 495 | 495 |
| Group | 46 | 45 | 45 | 45 | 45 |
| Wald stat | 559.6 | 262.6 | 1684.8 | 4269.0 | 360.2 |

Note: Year dummies are included in all estimations although its results are not reported. Numbers in parentheses are robust standard errors clustered at prefecture. *, **, and *** indicate significance at the 10%, 5%, and 1% levels.

Table 4. Determinants of number of entrants in Sumo stables for the period 1966-1985: Model corresponds to Table 3.

|  | (1) Poisson | (2) FE Pois | (3) FE Pois | (4) FE Pois |
|---|---|---|---|---|
| *Bad weather* | 0.03 (0.10) | −0.21** (0.08) | −0.12 (0.13) | −0.21** (0.10) |
| *RatEntrants 1868_45* | −0.04*** (0.01) | | | |
| *Bad weather* RatEntrants 1868_45* | | | −0.002 (0.008) | 0.001 (0.008) |
| *Bad weather* Agriculture* | | | −0.01 (0.01) | |
| *Distance* | 0.44*** (0.11) | | | |
| *Agriculture* | 0.07*** (0.01) | −0.004 (0.03) | −0.003 (0.03) | −0.005 (0.03) |
| *Population* | 0.42*** (0.02) | 0.58*** (0.13) | 0.57*** (0.14) | 0.58*** (0.13) |
| *GDP* | −0.41 (0.09) | 0.14 (0.14) | 0.12 (0.16) | 0.13 (0.17) |
| Observations | 920 | 900 | 900 | 900 |
| Group | 46 | 45 | 45 | 45 |
| Wald stat | 125.1 | 359.3 | 455.5 | 455.2 |

Note: Year dummies are included in all estimations although its results are not reported. Numbers in parentheses are robust standard errors clustered at prefecture. *, **, and *** indicate significance at the 10%, 5%, and 1% levels.

Table 5. Determinants of number of entrants in Sumo stables for the period 1966-1985:
  Number of new entrants in the pre-war period is replaced by their numbers in the post war period (1946-1965).

|  | (1) Poisson | (2) FE Pois | (3) FE Pois |
|---|---|---|---|
| *Bad weather* | −0.17* | 0.08 | −0.13 |
|  | (0.98) | (0.12) | (0.12) |
| *Rat Entrants 1946_65* | 0.11*** |  |  |
|  | (0.01) |  |  |
| *Bad weather\* Rat Entrants 1946_65* |  | −0.02 | −0.02 |
|  |  | (0.01) | (0.01) |
| *Bad weather\* Agriculture* |  | −0.01 |  |
|  |  | (0.01) |  |
| *Distance* | −0.04 |  |  |
|  | (0.14) |  |  |
| *Agriculture* | 0.03*** | −0.002 |  |
|  | (0.01) | (0.03) |  |
| *Population* | 0.24*** | 0.55*** | 0.56*** |
|  | (0.02) | (0.14) | (0.09) |
| *GDP* | −0.31*** | 0.11 | 0.12* |
|  | (0.08) | (0.14) | (0.07) |
| Observations | 920 | 900 | 900 |
| Group | 46 | 45 | 45 |
| Wald stat | 1430 | 348.5 | 341.4 |

Note: Year dummies are included in all estimations although its results are not reported. Numbers in parentheses are robust standard errors clustered at prefecture. *, **, and *** indicate significance at the 10%, 5%, and 1% levels.